\documentclass[twocolumn,showpacs,preprintnumbers,amsmath,amssymb]{revtex4}

\usepackage{graphicx}
\usepackage{dcolumn} 
\usepackage{bm}      

\usepackage{subfigure}
\usepackage[T1]{fontenc}
\usepackage{ae,aecompl}

\begin{document}

\title{Langevin Dynamics in Constant Pressure Extended Systems}

\author{D Quigley}
\email{dq100@york.ac.uk}
\author{MIJ Probert}
\email{mijp1@york.ac.uk}
\affiliation{
University of York\\
Heslington \\
York \\
United Kingdom \\
YO10 5DD 
}

\date{\today}

\begin{abstract}
The advantages of performing Langevin Dynamics in extended systems are
discussed. A simple Langevin Dynamics scheme for producing the canonical
ensemble is reviewed, and is then extended to the Hoover ensemble. We
show that the resulting equations of motion generate the isobaric-isothermal
ensemble. The Parrinello-Rahman ensemble is then discussed and we show
that despite the presence of intrinsic probability gradients in this system, 
a Langevin Dynamics approach samples the extended phase space in the correct
fashion. The implementation of these methods in the \emph{ab-initio} plane wave
density functional theory (DFT) code CASTEP 
[M. D. Segall, P. L. D. Lindan, M. J. Probert, C. J. Pickard, P. J. Hasnip, S. J.
Clarke and M.C. Payne, J. Phys.: Cond. Matt. {\bf 14} (11), 2717 (2003)]
is demonstrated.
\end{abstract}


\maketitle


\section{Introduction}

The method of molecular dynamics is a well established
tool with wide ranging applications. Of particular
advantage is the ability to obtain ensemble
averages of statistical quantities (in ergodic systems)
and detailed trajectory information within the same
computational framework.

Many of the early attempts to extend molecular
dynamics from the micro-canonical (NVE)
to the canonical (NVT) ensemble used
some form of stochastic dynamics. A random
component to the particle dynamics stimulates 
ergodicity and can reduce correlations
for increased sampling efficiency at the expense
of accuracy in short-term dynamics. A simple
stochastic dynamics scheme based on the
Langevin equation will be reviewed in
section \ref{sec:NVTLangevin}.

A more widely adopted scheme for generating
NVT ensemble trajectories is the Nos\'e
thermostat \cite{Nose84a,Hoover85}. Here
time is rescaled according to an extra `extended system'
variable. This is introduced to the Hamiltonian
in a manner chosen to reproduce the correct
NVT partition function and hence simulate
coupling to a heat bath. Use of this
method requires either a non-uniform
sampling in the scaled time variable \cite{Nose84a} 
or a non-canonical
transform to un-scaled time \cite{Hoover85}. Recent
work on non-Hamiltonian statistical mechanics
has shown that this is not a major concern 
for the trajectory of the particle subsystem
\cite{TuckermanMM99,TuckermanLCM01}. Other
work has reformulated the Nos\'e scheme to remove
the need for a non-canonical transform \cite{BondLL99}
altogether.

Of greater concern is the problem of ergodicity. In
order to generate correct ensemble averages from
MD trajectories, the simulated system \emph{must}
be ergodic. The simple deterministic nature
of the Nos\'e scheme requires that the chaotic
behaviour needed to stimulate ergodicity is
provided by the particle sub-system itself. For
certain systems, in particular those containing
harmonic forces, this requirement is known not to
be satisfied \cite{Hoover85} leading to incorrect phase-space
trajectories and hence poor ensemble averages. It is now common
practise to overcome this problem by the use of a chain of Nos\'e-style
thermostats \cite{MartynaKT92}. Although guaranteed to
correctly sample the isothermal ensemble, this scheme
is still deterministic and exhibit longer
correlation times than stochastic methods in some circumstances.
Note that recent work has generalised the Nos\'e scheme
to include more chaotic terms in the extended Lagrangian \cite{GenNH,LLGenNH} which
does much to eliminate these issues.

Extended Hamiltonian methods are also available to
simulate the coupling of the particle system to 
a pressure piston. This idea was introduced
by Andersen \cite{Andersen80} in a scheme where
particle positions and momenta are scaled according
to the cell volume. Potential and kinetic energy
terms for the volume are added to the Lagrangian
resulting in an expansion or contraction of the 
system to regulate pressure with the required
fluctuations. A modification of the Andersen scheme
in which the extended system variables are the
strain rate $\dot{.psilon}$ and its conjugate, was
introduced by Hoover \cite{Hoover86}. This
tends to be more robust in practise. A scheme
in which the cell motion is anisotropic with each 
cell vector evolving independently was introduced
by Parrinello and Rahman  \cite{ParRahman80,ParrinelloR81}.

These constant pressure schemes have been 
coupled to Nos\'e style thermostats resulting in
schemes for sampling the isobaric-isothermal (NPT)
ensemble \cite{MTKCorrection,SturgeonL00}. 

Kolb and Dunweg \cite{KolbD99} have shown that an
effective method for sampling the isobaric-isothermal
ensemble can be obtained by performing Langevin dynamics
in the extended Andersen system. This has the potential
advantage of shorter correlation times and guaranteed ergodicity in the NPT
ensemble. In this paper, we show that Langevin dynamics
can be conducted within a Hoover-style extended
system to correctly sample the isothermal-isobaric
ensemble. The extension of these ideas to Langevin
dynamics in a Parrinello-Rahman system will
also be discussed. The resulting equations of motion are
integrated using an evolution algorithm 
obtained from Louiville time evolution operators \cite{TuckermanBM92}
which is symplectic in the limit of zero friction.

Criteria for selecting coupling parameters for
the thermostatic/barostatic processes will also be discussed and
applied to various example simulations.


\section{Langevin Dynamics}
\label{sec:NVTLangevin}

A simple but effective method of performing Langevin dynamics simulations
for the isothermal ensemble uses the following equations of motion in the
usual notation

\begin{subequations}
\begin{eqnarray}
\label{eq:La}
\dot{\mathbf{r}}_{i} &=& \mathbf{p}_{i}/m_{i} \\
\label{eq:Lb}
\dot{\mathbf{p}}_{i} &=&  \mathbf{f}_{i} - {\gamma}\mathbf{p}_{i} + \mathbf{R}_{i}.
\end{eqnarray}
\end{subequations}

Here $\gamma$ is a friction co-efficient representing viscous damping
due to fictitious `heat bath' particles. The random force
$\mathbf{R}_{i}$ represents the effect of collisions with these
particles. Following Chandrasekhar \cite{Chand43} we assume
that the timescale of the collisional heat bath process is very
much smaller than the ionic motions of interest, and hence

\begin{equation}
\left<
\mathbf{R}_{i}\left(t\right)
\mathbf{R}_{i}\left(t^{\prime}\right)
\right>
=\delta\left(t-t^{\prime}\right).
\end{equation}

Furthermore, we assume that a great many collisions with
heat bath particles occur over a MD time-step and we can
therefore expect that $\mathbf{R}_{i}$ will be distributed
in a Gaussian fashion in accordance with the central limit
theorem. The Stokes-Einstein relation for the diffusion co-efficient
can then be used to show that the average value of $\mathbf{R}_{i}$
over a time-step (in thermal equilibrium) should be a random
deviate drawn from a Gaussian distribution of zero mean and
unit variance scaled by

\begin{equation}
\label{eq:balance}
\sqrt{\frac{2k_{B}T\gamma m }{\Delta t}}.
\end{equation}

Note that this choice of $\mathbf{R}_{i}$ limits this simple scheme, and
the constant pressure schemes that follow, to equilibrium
simulations only. Certain deterministic thermostats can however be
utilised in non-equilibrium simulations \cite{Hoover2004}.

It can be shown that equations \ref{eq:La} and \ref{eq:Lb} are
equivalent to the following Fokker-Planck equation for the phase space
probability density $\rho\left(\mathbf{r}^{N},\mathbf{p}^{N}\right)$

\begin{eqnarray}
\label{eq:NVTFokkerPlanck}
\frac{\partial}{\partial{t}}\rho
+\sum_{i=1}^{N}\left[
\frac{\mathbf{p}_{i}}{m}\cdot\nabla_{\mathbf{r}_{i}}
\rho
+\mathbf{f}_{i}\cdot\nabla_{\mathbf{p}_{i}}
\rho
\right] \nonumber \\ 
= \gamma\sum_{i=1}^{N}\nabla_{\mathbf{p}_{i}}
\cdot\left[\mathbf{p}_{i}\rho
+mk_{B}T\nabla_{\mathbf{p}_{i}}
\rho\right].
\end{eqnarray}

This has the canonical phase space probability
density function $\rho_{_{NVT}}$ as a stationary
solution, and hence the method can be used to 
sample the isothermal ensemble via a fluctuation-
dissipation process\cite{AllenTildesley}.

The choice of the parameter $\gamma$ is a compromise between 
statistical sampling efficiency and preservation of accuracy
in short-term dynamics. However, in the case where the process
approximated by the stochastic components can be simulated by another method,
it is possible to determine an optimal value of $\gamma$
numerically. Guidelines for the choice of $\gamma$
will be discussed in detail later. 

An effective Verlet-style integrator can be obtained for this
scheme by applying the time-evolution operator formulation of Tuckerman
\emph{et al} \cite{TuckermanBM92} in the limit $\gamma\rightarrow 0$. The lack
of a Louiville operator for the stochastic components of the dynamics requires
that these are then included into the particle forces via the following
substitutions:

\begin{eqnarray}
\mathbf{f}_{i}\left(t\right)
&\rightarrow&
\mathbf{f}_{i}\left(t\right) 
- \gamma \mathbf{p}_{i}\left(t\right)
+ \mathbf{R}_{i}\left(t\right) \nonumber \\
\mathbf{f}_{i}\left(t+\Delta t\right)
&\rightarrow&
\mathbf{f}_{i}\left(t+\Delta t\right)
- \gamma \mathbf{p}_{i}\left(t +\Delta t\right)
+ \mathbf{R}_{i}\left(t+\Delta t\right) \nonumber
\end{eqnarray}

In the NVT case described here, this leads to standard
Velocity-Verlet, modified with the above substitutions.


\section{Langevin-Hoover Dynamics}

In this section, we show that the simple NVT
scheme above can be extended to perform Langevin
dynamics in a Hoover-style extended system, and that 
this results in correct sampling of the isobaric-isothermal
ensemble.

\subsection{Equations of Motion}

The equations of motion we propose are shown below. These 
incorporate the constant pressure component of the Hoover
equations as corrected by Martyna \emph{et al} \cite{MTKCorrection}.
In the form shown below, the deterministic equations for the
evolution of both the particle and barostat velocity have
been converted to Langevin equations in $d$ dimensions with different 
friction constants

\begin{subequations}
\begin{eqnarray}
\label{eq:LHa}
\dot{\mathbf{r}}_{i}&=&\frac{\mathbf{p}_{i}}{m_{i}}
+ \frac{p_{.psilon}}{W}\mathbf{r}_{i}\\
\label{eq:LHb}
\dot{\mathbf{p}}_{i}&=& -\nabla_{\mathbf{r}_{i}}\Phi
-\left(1+\frac{d}{N_{f}}\right)\frac{p_{.psilon}}{W}\mathbf{p}_{i}
-\gamma\mathbf{p}_{i}+\mathbf{R}_{i}  \\
\label{eq:LHc}
\dot{\mathcal{V}} &=& d\mathcal{V}p_{.psilon}/W  \\
\label{eq:LHd}
\dot{p}_{.psilon} &=& d\mathcal{V}\left(X-P_{ext}\right)
+\frac{d}{N_{f}}\sum_{i=1}^{N}\frac{\mathbf{p}^{2}_{i}}{m_{i}} 
-\gamma_{p}p_{.psilon}+R_{p}.
\end{eqnarray}
\end{subequations}

These equations introduce the volume $\mathcal{V}$ as a dynamical (barostat)
variable. The corresponding momentum variable $p_{.psilon}$ is
the strain rate $\dot{.psilon}$ multiplied by a fictitious mass $W$.
$R_{p}$ is a stochastic `force' which acts on the barostat. The use
of a Langevin equation for the barostat as well as the particles
may have possible equilibration benefits, but we shall see in the analysis
that follows that it is not required for canonical sampling at equilibrium.

The scalar $X$ is given by

\begin{equation}
X= \frac{1}{d\mathcal{V}}\left[\sum_{i=1}^{N}
\frac{\mathbf{p}_{i}\cdot\mathbf{p}_{i}}{m_{i}}
+ \sum_{i=1}^{N}\mathbf{r}_{i}\cdot\mathbf{f}_{i}
\right] - \frac{\partial}
{\partial\mathcal{V}} \Phi\left(\mathbf{r}^{N},\mathcal{V}\right).
\end{equation}

The distinction between $X$ and the instantaneous pressure
$\mathcal{P}$ is important. The value of $\mathcal{P}$ when
calculated using the virial equation should include the
white noise contributions from the `Langevin thermostat' whereas
$X$ does not. Use of $\mathcal{P}$ in equation \ref{eq:LHd}
leads to non-canonical temperature and volume fluctuations. 

The values of $\mathbf{R}_{i}$ are drawn from the same distribution
as the NVT case. Values of $R_{p}$ are drawn from a Gaussian
distribution of zero mean and unit variance scaled by

\begin{equation}
\sqrt{\frac{2k_{B}TW\gamma_{p} }{\Delta t}}.
\end{equation}

In the un-thermostatted limit ($\gamma \rightarrow 0, \gamma_{p}
\rightarrow 0$) equations \ref{eq:LHa} to \ref {eq:LHb} obey the
Louiville theorem 

\begin{equation}
\frac{\partial \rho}{\partial t}
+\sum_{i=1}^{N}\mathbf{\dot{r}}_{i}\nabla_{\mathbf{r}_{i}}\rho
+\sum_{i=1}^{N}\mathbf{\dot{p}}_{i}\nabla_{\mathbf{p}_{i}}\rho
+\dot{p}_{.psilon}\frac{\partial \rho}{\partial p_{\epsilon}}
+\dot{\mathcal{V}}\frac{\partial \rho}{\partial \mathcal{V}} = 0
\end{equation}
and conserve the quantity

\begin{equation}
H^{\prime}=\mathcal{H}\left(\mathbf{r}^{N},\mathbf{p}^{N}\right) + P\mathcal{V}
  + p^{2}_{.psilon}/2W,
\end{equation}
where $\mathcal {H}$ is the Hamiltonian of the particle sub-system.
Note that $H^{\prime}$ itself is not a Hamiltonian. A rigorous phase space analysis
is therefore best performed using techniques in non-Hamiltonian
statistical mechanics as introduced by Tuckerman and co-workers
\cite{TuckermanMM99,TuckermanLCM01}.


\subsection{Justification}

To show that equations \ref{eq:LHa} to \ref{eq:LHd}
correctly sample the isobaric-isothermal ensemble
we first identify the un-thermostatted phase-space probability 
density for the extended system of particles,

\begin{equation}
\label{eq:NPHExtPhaseDensity}
\rho_{{}_{NPH^{\prime}}}\left(\mathbf{r}^{N},\mathbf{p}^{N},p_{.psilon},\mathcal{V}\right)
\propto  
\frac{\delta\left[H^{\prime}(t)- H^{\prime}(0)\right)}{\Omega_{NPH^{\prime} }
\left(\mathbf{r}^{N},\mathbf{p}^{N},p_{.psilon},\mathcal{V}\right]}.
\end{equation}

The phase space density for this system coupled to a heat bath
at constant temperature should therefore be

\begin{eqnarray}
\label{eq:NPTExtPhaseDensity}
\rho_{{}_{NPT}}\left(\mathbf{r}^{N},\mathbf{p}^{N},p_{.psilon},\mathcal{V}\right) \propto  
\frac{1}{\Omega_{NPT}\left(\mathbf{r}^{N},\mathbf{p}^{N},p_{.psilon},\mathcal{V}\right)}
\nonumber \\ \times 
\exp{\left[-\frac{\mathcal{H}\left(\mathbf{r}^{N},\mathbf{p}^{N}\right)
+ P\mathcal{V}
  + p^{2}_{.psilon}/2W}
  {k_{B}T}\right]}.
\end{eqnarray}

This is the probability
density function for the correctly thermostatted 
extended particle plus barostat phase space. Integration over the
barostat momentum $p_{.psilon}$ yields a constant, and hence

\begin{eqnarray}
\label{eq:NPTPhaseDensity}
\rho_{{}_{NPT}}\left(\mathbf{r}^{N},\mathbf{p}^{N},\mathcal{V}\right) \propto  
\frac{1}{\Omega_{NPT}\left(\mathbf{r}^{N},\mathbf{p}^{N},\mathcal{V}\right)}
\nonumber \\ \times 
\exp{\left[-\frac{\mathcal{H}\left(\mathbf{r}^{N},\mathbf{p}^{N}\right) + 
P\mathcal{V}}
  {k_{B}T}\right]}.
\end{eqnarray}
which is the correct probability density
function for the isobaric-isothermal
particle subsystem.

Following Stratonovich \cite{Stratonovich} we now
construct the following Fokker-Planck equation for the phase space
density $\rho$ resulting from equations \ref{eq:LHa} to \ref{eq:LHd}. 
The compressibility of equations \ref{eq:LHa} to \ref{eq:LHd} is zero
and hence the Jacobian of the resulting co-ordinate transform is
the identity matrix and we need not include it:

\begin{widetext}
\begin{eqnarray}
\label{eq:NPTFokkerPlanck}
\frac{\partial\rho}{\partial{t}}&+&
\sum_{i=1}^{N}\left\{
\left(\frac{\mathbf{p}_{i}}{m_{i}} + \frac{p_{.psilon}}{W}\mathbf{r}_{i}\right)
\cdot\nabla_{\mathbf{r}_{i}}
\rho
+\left[\mathbf{f}_{i}-\left(1+\frac{d}{N_{f}}\right)\frac{p_{.psilon}}{W}\mathbf{p}_{i}\right]
\cdot\nabla_{\mathbf{p}_{i}}
\rho
\right\} \nonumber\\
&+& \left[d\mathcal{V}\left(\chi-P_{ext}\right)
+\frac{d}{N_{f}}\sum_{i=1}^{N}\frac{\mathbf{p}^{2}_{i}}{m_{i}} 
\right]\frac{\partial\rho}{\partial p_{.psilon}}
+ \dot{\mathcal{V}}\frac{\partial\rho}{\partial\mathcal{V}} \nonumber \\
&=& \gamma_{p}\frac{\partial}{{\partial}{p_{.psilon}}}
\left[p_{.psilon}\rho
+Wk_{B}T\frac{\partial}{{\partial}{p_{.psilon}}}
\rho\right]
+ \gamma\sum_{i=1}^{N}\nabla_{\mathbf{p}_{i}}
\cdot\left[\mathbf{p}_{i}\rho
+mk_{B}T\nabla_{\mathbf{p}_{i}}
\rho\right].
\end{eqnarray}
\end{widetext}

where $\rho = \rho\left(\mathbf{r}^{N},\mathbf{p}^{N},p_{.psilon},
\mathcal{V}\right)$.

In order for the Langevin-Hoover scheme to correctly
sample the isobaric-isothermal ensemble, equation \ref{eq:NPTExtPhaseDensity} must
be a solution of \ref{eq:NPTFokkerPlanck}. As the un-thermostatted system
is Louivillian, we expect the LHS of equation  \ref{eq:NPTFokkerPlanck} to
be identically zero which is easily confirmed.Our use of the Stokes-Einstein
relation in balancing the stochastic part of the 
Langevin dynamics with the dissipative friction term ensures that
the RHS is also zero for any phase space probability density function
representing equilibrium with a heat bath. This is also easily confirmed
in the case where $\rho$ is given by equation \ref{eq:NPTExtPhaseDensity}, and
hence the proposed equations of motion are expected to sample the
isobaric-isothermal ensemble in the correct fashion.

A complete justification that the Langevin-Hoover scheme samples the
required ensemble must show that the desired temperature and pressure
are maintained with canonical fluctuations about the mean of each. This
behaviour depends on the choice of the parameters $W$, $\gamma$ and $\gamma_{p}$,
and will be covered in detail in section \ref{subsec:choice}.


\subsection{Numerical Integration}
\label{subsec:numint}

In this section we follow the method discussed above
for obtaining integration algorithms for systems
obeying Langevin equations. The
Louiville operator for the isotropic Langevin NPT equations of motion
in the limit of zero friction is

\begin{eqnarray}
iL&=&\dot{\mathbf{r}}\frac{\partial}{\partial \mathbf{r}} +
\dot{\mathbf{p}}\frac{\partial}{\partial \mathbf{p}} +
\dot{.psilon}\frac{\partial}{\partial \epsilon} +
\dot{p_{.psilon}}\frac{\partial}{\partial p_{\epsilon}} \nonumber \\
&=&iL_{r} + iL_{p} + iL_{.psilon} + iL_{p_{\epsilon}}.
\end{eqnarray}

The following Trotter factorisation of the resulting time-step
evolution operators was found to be the most
convenient:

\begin{multline}
e^{iL\Delta t} = e^{iL_{.psilon}\Delta t/2}
e^{iL_{p_{.psilon}}\Delta t/2}
e^{iL_{p}\Delta t/2}
e^{iL_{r}\Delta t} \times  \\
e^{iL_{p}\Delta t/2} 
e^{iL_{p_{.psilon}}\Delta t/2}
e^{iL_{.psilon}\Delta t/2}
\end{multline}
which leads to the following integration algorithm.

\begin{enumerate}
\item{
$\mathcal{V}^{t+\frac{1}{2}\Delta t} =\mathcal{V}^{t}
+\frac{\Delta t}{2}\dot{\mathcal{V}}\left[
\mathcal{V}^{t},p_{.psilon}^{t}
\right]$}
\item{
$p_{.psilon}^{t+\frac{1}{2}\Delta t} = p_{\epsilon}^{t}
+\frac{\Delta t}{2}\dot{p}_{.psilon}\left[
\mathbf{r}_{i}^{t},\mathbf{p}_{i}^{t},
\mathcal{V}^{t+\frac{1}{2}\Delta t},p_{.psilon}^{t}
\right]$}
\item{
$\mathbf{p}_{i}^{t+\frac{1}{2}\Delta t} = \mathbf{p}_{i}^{t}
+\frac{\Delta t}{2}\dot{\mathbf{p}}_{i}\left[
\mathbf{r}_{i}^{t},\mathbf{p}_{i}^{t},
p_{.psilon}^{t+\frac{1}{2}\Delta t}
\right]$}
\item{
$\mathbf{r}_{i}^{t+\Delta t} = \mathbf{r}_{i}^{t}
+\Delta t\dot{\mathbf{r}}_{i}\left[
\mathbf{r}_{i}^{t},\mathbf{p}_{i}^{t+\frac{1}{2}\Delta t},
p_{.psilon}^{t+\frac{1}{2}\Delta t}
\right]$}
\item{
$\mathbf{p}_{i}^{t+\Delta t} = \mathbf{p}_{i}^{t+\frac{1}{2}\Delta t}
+\frac{\Delta t}{2}\dot{\mathbf{p}}_{i}\left[
\mathbf{r}_{i}^{t+\Delta t},\mathbf{p}_{i}^{t+\frac{1}{2}\Delta t},
p_{.psilon}^{t+\frac{1}{2}\Delta t}
\right]$}
\item{
$p_{.psilon}^{t+\Delta t} = p_{\epsilon}^{t+\frac{1}{2}\Delta t}$
\begin{flushright}
$+\frac{\Delta t}{2}\dot{p}_{.psilon}\left[
\mathbf{r}_{i}^{t+\Delta t},\mathbf{p}_{i}^{t+\Delta t},
\mathcal{V}^{t+\frac{1}{2}\Delta t},p_{.psilon}^{t+\frac{1}{2}\Delta t}
\right]
$
\end{flushright}
}
\item{
$\mathcal{V}^{t+\Delta t} = \mathcal{V}^{t}
+\frac{\Delta t}{2}\dot{\mathcal{V}}\left[
\mathcal{V}^{t+\frac{1}{2}\Delta t},p_{.psilon}^{t+\Delta t}
\right].  
$}
\end{enumerate}
Making the appropriate substitution for the particle forces
at each time-step we now include the Langevin buffeting and damping
terms and denote these as extra dependences of the time derivatives.

\begin{enumerate}
\item{$
\mathcal{V}^{t+\frac{1}{2}\Delta t} =\mathcal{V}^{t}
+\frac{\Delta t}{2}\dot{\mathcal{V}}\left[
\mathcal{V}^{t},p_{.psilon}^{t}
\right]$} 
\item{$
p_{.psilon}^{t+\frac{1}{2}\Delta t} = p_{\epsilon}^{t}
+\frac{\Delta t}{2}\dot{p}_{.psilon}\left[
\mathbf{r}_{i}^{t},\mathbf{p}_{i}^{t},
\mathcal{V}^{t+\frac{1}{2}\Delta t},p_{.psilon}^{t},
\gamma_{p}p_{.psilon}^{t},R_{p}^{t}
\right]
$}
\item{$
\mathbf{p}_{i}^{t+\frac{1}{2}\Delta t} = \mathbf{p}_{i}^{t}
+\frac{\Delta t}{2}\dot{\mathbf{p}}_{i}\left[
\mathbf{r}_{i}^{t},\mathbf{p}_{i}^{t},
p_{.psilon}^{t+\frac{1}{2}\Delta t},
\gamma \mathbf{p}_{i}^{t},\mathbf{R}_{i}^{t}
\right]
$}
\item{$
\mathbf{r}_{i}^{t+\Delta t} = \mathbf{r}_{i}^{t}
+\Delta t\dot{\mathbf{r}}_{i}\left[
\mathbf{r}_{i}^{t},\mathbf{p}_{i}^{t+\frac{1}{2}\Delta t},
p_{.psilon}^{t+\frac{1}{2}\Delta t}
\right]
$}
\item{$
\mathbf{p}_{i}^{t+\Delta t} = \mathbf{p}_{i}^{t+\frac{1}{2}\Delta t}
$
\\ \begin{flushright}
$
+\frac{\Delta t}{2}\dot{\mathbf{p}}_{i}\left[
\mathbf{r}_{i}^{t+\Delta t},\mathbf{p}_{i}^{t+\frac{1}{2}\Delta t},
p_{.psilon}^{t+\frac{1}{2}\Delta t}
\gamma \mathbf{p}_{i}^{t+\Delta t},\mathbf{R}_{i}^{t+\Delta t}
\right]
$
\end{flushright}
}
\item{$
p_{.psilon}^{t+\Delta t} = p_{\epsilon}^{t+\frac{1}{2}\Delta t}
+$
\\ \begin{flushright}
$\frac{\Delta t}{2}\dot{p}_{.psilon}\left[
\mathbf{r}_{i}^{t+\Delta t},\mathbf{p}_{i}^{t+\Delta t},
\mathcal{V}^{t+\frac{1}{2}\Delta t},p_{.psilon}^{t+\Delta t},
\gamma_{p}p_{.psilon}^{t+\Delta t},R_{p}^{t+\Delta t}
\right]
$
\end{flushright}
}
\item{$
\mathcal{V}^{t+\Delta t} = \mathcal{V}^{t}
+\frac{\Delta t}{2}\dot{\mathcal{V}}\left[
\mathcal{V}^{t+\frac{1}{2}\Delta t},p_{.psilon}^{t+\Delta t}
\right] $.}
\end{enumerate}

To demonstrate the stability of this algorithm we consider
the limit of vanishing friction in which $\gamma \rightarrow 0$.
In this limit the system is Louivillian and the quantity

\begin{equation}
H^{\prime} = \Phi \left(\{\mathbf{r}_{i}\}\right) +
\sum_{i=1}^{N}\mathbf{p}^{2}_{i}/2m_{i}
+ p_{.psilon}^{2}/2W + P_{ext}\mathcal{V}
\end{equation}
is conserved. Figure 1 shows the value of
$H^{\prime}$ 
for a Lennard-Jones system both in the zero friction limit, and for
a Langevin-Hoover simulation.
With the introduction of finite
friction we expect deviations from $H^{\prime}$ with
no long term drift as indicated.

\begin{figure}
\includegraphics*[scale=0.5]{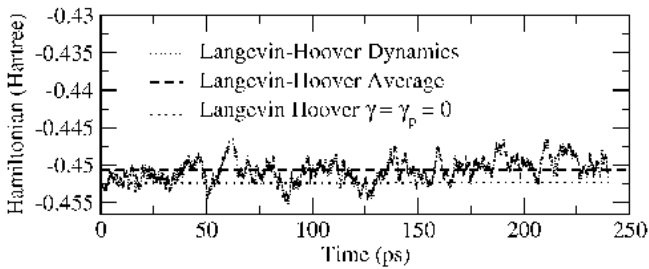}
\caption{Value of the conserved quantity in the Hoover system for a Langevin dynamics
simulation, and a similar simulation in the zero friction limit. Interactions
were modelled with the Lennard-Jones potential for argon at, 67.6\,MPa. In the case
of the Langevin dynamics simulation, a set temperature of 20\,K was used. An equilibration
time of 10,000 $\Delta t$ was used prior to sampling. The drift in the zero friction
value is less than $2 \times 10^{-4}$ Hartree over the 100,000 time-step run with 
$\Delta t = 2.4$\,fs.}
\label{fig:nofrictionstability}
\end{figure}


\subsection{Choice of Parameters}
\label{subsec:choice}

The isotropic scheme uses three parameters which must be
chosen carefully. These are the barostat `mass' $W$, the
particle friction coefficient $\gamma$ and the 
barostat friction coefficient $\gamma_{p}$.

It has been previously noted \cite{MartynaTTK96} that for a Hoover barostat, the
fictitious mass should be chosen according to

\begin{equation}
W = dNk_{b}T/ \omega_{b}^{2}
\end{equation}
where $\omega_{b}$ is the frequency of the required volume
fluctuations. The choice of this $\omega_{b}$ depends on the 
timescale on which the particle motions of interest occur.
A useful aide when choosing this frequency is the NVT
temperature spectrum (TS). The character of this spectrum
should not be significantly disrupted by the addition of a barostat. In practise
this means choosing $\omega_{b}$ to be less than the
smallest frequency component in the TS (figure 2). Care
must also be taken that $\omega_{b}$ is sufficiently close to these components that
de-coupling of the barostat from the particle subsystem is avoided.

\begin{figure}
\includegraphics*[scale=0.4]{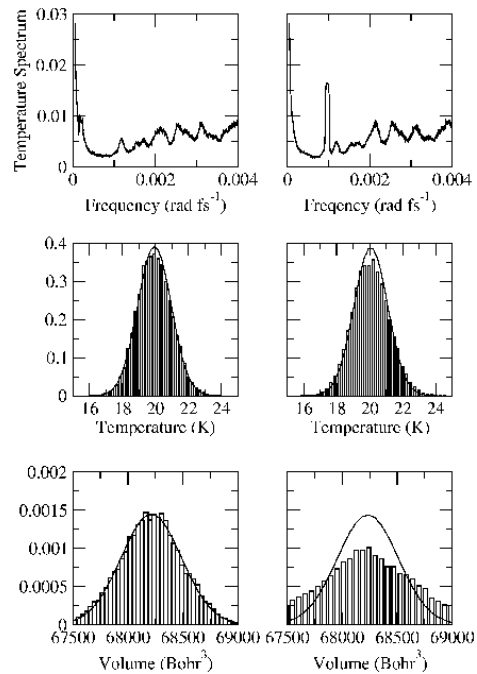}
\caption{Choice of $\omega_{b}$ for solid Lennard-Jones argon at
50\,K 20\,MPa. A thermostat frequency $\omega_{p}=2\pi\gamma=6.6\times 10^{-4}$\,
rad fs$^{-1}$ is used in both cases. In the left hand case, $\omega_{b}$
is chosen to be less than the lowest frequency which appears in the NVT
temperature spectrum. The corresponding temperature and volume
sample distributions match the exact canonical (solid line) case
exactly. In the right hand case the choice of $\omega_{b}$ has disrupted
the natural temperature spectrum resulting in non-canonical
fluctuations.}
\label{fig:wbsolid1}
\end{figure}

To determine an appropriate value for the particle friction
coefficient $\gamma$ we calculate the memory function for the
system we wish to study. The memory function $\xi$ for
the velocity autocorrelation function $\psi$ is defined by

\begin{equation}
\frac{d\psi}{dt}=
-\int^{t}_{0}\xi\left(t-\tau\right)\psi\left(\tau\right)
d\tau .
\end{equation}

The generalised Langevin equation is

\begin{equation}
\dot{\mathbf{p}}_{i}= \mathbf{F}_{i} 
-\int^{t}_{0}\xi\left(t-\tau\right)\mathbf{p}_{i}\left(\tau\right)d\tau
+\mathbf{R}_{i},
\end{equation}
which in the extended particle + barostat
phase space becomes


\begin{multline}
\dot{\mathbf{p}}_{i}= -\nabla_{\mathbf{r}_{i}}\Phi 
-\left(1+\frac{d}{N_{f}}\right)\frac{p_{.psilon}}{W}\mathbf{p}_{i} \\
-\int^{t}_{0}\xi\left(t-\tau\right)\mathbf{p}_{i}\left(\tau\right)
+\mathbf{R}_{i}.
\end{multline}

Comparing to equation \ref{eq:LHb}, we can see that the stochastic component
of the dynamics is generated within the approximation

\begin{equation}
\xi = \gamma \delta(t).
\end{equation}

To retain consistency with this approximation, the value of gamma
should be equal to

\begin{equation}
\gamma = \xi_{0} = \int_{0}^{\infty} \xi_{act}(\tau) d\tau ,
\end{equation}
where $\xi_{act}$ is the actual memory function of the system we wish
to simulate. Calculation of the optimal $\gamma$ therefore requires
prior knowledge of the memory function which can be obtained from
an NVE simulation or from other NVT/NPT techniques. As the stochastic
component of the dynamics has now been chosen to be characteristic
of the particle motions themselves, the `heat bath' 
is now representative  of the effect of the bulk on our simulated 
sample. This is particularly appropriate for simulations using
periodic boundary conditions, where the only physical objects
with which the simulated particles exchange heat, are other particles
in neighbouring cells.

A useful method of calculating $\xi (t)$ from trajectory information
via an autoregressive model has been proposed by Kneller and Hinsen
\cite{KnellerH01}. We will use this method to calculate
appropriate value of $\gamma$ for various simple systems in
section \ref{sec:Examples}.

The cost of computing a memory
function in advance of a Langevin dynamics simulation can 
sometimes be prohibitive and it is often convenient to adopt the
conservative rule of thumb that the thermostat frequency $\omega_{p}=
2\pi/\gamma$ should be a few
tens of times smaller than the smallest characteristic frequency
of the system to be studied. This may result in less than
optimum sampling efficiency, but will generate the NPT
ensemble without disrupting the particle motions of interest.

The choice of the `piston thermostat' parameter $\gamma_{p}$ is
of less importance. The inclusion of the damping and buffeting
terms in equation \ref{eq:LHd} is not necessary for reproduction of
the NPT partition function, however a non-zero $\gamma_{p}$
may equilibration times. Furthermore the effect on the
particle velocities is small in equilibrium. We generally require
that $2\pi/\gamma_{p}$ be ten times smaller than the frequency
associated with the barostat variable. It is in principle
possible to calculate an autocorrelation function
for $p_{.psilon}$ and hence a memory function and optimal
barostatic friction coefficient. This is not useful in 
practise.


\section{Langevin-Parrinello-Rahman Dynamics}

\subsection{Equations of Motion}

An implementation of Langevin Dynamics in a fully flexible
simulations cell follows from the
Nos\'e-Hoover thermostatted Parrinello-Rahman scheme 
of Martyna, Tobias and Klein \cite{MTKCorrection} after
removing the Nos\'e-Hoover chains and converting
to Langevin equations in the momenta:

\begin{widetext}
\begin{subequations}
\begin{eqnarray}
\label{eq:LPRa}
\dot{\mathbf{r}}_{i} &=&  \frac{\mathbf{p}_{i}}{{m}_{i}} +
\frac{\mathbf{p}_{g}}{W_{g}}\mathbf{r}_{i}  \\
\label{eq:LPRb}
\dot{\mathbf{p}}_{i} &=& -\nabla_{\mathbf{r}_{i}}\Phi\left(\mathbf{r}^{N}
,\mathbf{h}\right)
-\frac{\mathbf{p}_{g}}{W_{g}}\mathbf{p}_{i}
-\left(\frac{1}{N_{f}}\right)
\frac{\mathrm{Tr}\left[\mathbf{p}_{g}\right]}{W_{g}}{\mathbf{p}}_{i}
-\gamma\mathbf{p}_{i} + \mathbf{R}_{i}  \\
\label{eq:LPRc}
\dot{\mathbf{h}}&=& \frac{\mathbf{p}_{g}\mathbf{h}}{W_{g}}  \\
\label{eq:LPRd}
\dot{\mathbf{p}}_{g} &=& \mathcal{V}\left(\mathbf{X}
-{P}_{ext}\mathbf{I}\right) + \left[\frac{1}{N_{f}}
\sum_{i=1}^{N}\frac{\mathbf{p}^{2}_{i}}{m_{i}}\right]\mathbf{I}
-\gamma_{p}\mathbf{p}_{g}+\mathbf{R}_{p}
\end{eqnarray}
\end{subequations}
\end{widetext}
where the tensor $\mathbf{X}$ is given by

\begin{eqnarray}
\label{eq:chitensor}
\mathbf{X}_{\alpha,\beta}&=& \frac{1}{\mathcal{V}}
\left[\sum_{i=1}^{N}
     \frac{\left(\mathbf{p}_{i}\right)_{\alpha}\left(\mathbf{p}_{i}\right)_{\beta}}
           {m_{i}}+{\left(\mathbf{r}_{i}\right)_{\alpha}\left(\mathbf{f}_{i}\right)_{\beta}}
-\left(
       \phi^{\prime} \mathbf{h}^{T}
\right)_{\alpha,\beta}
\right] \nonumber \\
\left(\phi^{\prime}\right)_{\alpha,\beta}&=&\frac{\partial \phi
\left(\mathbf{r}^{N},\mathbf{h}\right)}{\partial \left(\mathbf{h}\right)_{\alpha,\beta}}
\end{eqnarray}

As before, we draw the random forces $\mathbf{R}_{i}$ from the
same distribution as in the NVT case. Each component
of the barostat buffeting tensor $\mathbf{R}_{p}$ is drawn
from a Gaussian distribution on unit mean and zero
variance scaled by

\begin{equation}
\sqrt{\frac{2k_{B}W_{g}\gamma_{p} }{\Delta t}}.
\end{equation}
where $W_{g}$ is set by the barostatic frequency $\omega_{b}$
according to

\begin{equation}
W_{g}=(N_{f}+d)k_{B}T/d\omega_{b}^{2}.
\end{equation}

These equations can be evolved using an analagous integration
scheme to that described for the isotropic case in section \ref{subsec:numint}.
Approprate frequencies for the thermostat and barostat are also chosen in
the same way as for the isotropic scheme.


\subsection{Analysis}

Equations \ref{eq:LPRa} to \ref{eq:LPRd} conserve the quantity

\begin{equation}
\label{eq:PRHPrime}
H^{\prime}=\mathcal{H}\left(\mathbf{r}^{N},\mathbf{p}^{N}\right)
+\frac{1}{2W_{g}}\mathrm{Tr}\left[\mathbf{p}_{g}\mathbf{p}_{g}^{T}\right]
+P_{ext}\det\left[\mathbf{h}\right]
\end{equation}
in the limit of zero friction (the pure Parrinello-Rahman system). As 
in the Hoover case, this is not a true Hamiltonian. The system
of equations cannot be derived from equation \ref{eq:PRHPrime}
via Hamiltons equations. The phase space analysis therefore requires
the tools of Tuckerman \emph{et al} \cite{TuckermanMM99,TuckermanLCM01}, and
merits more attention than in the Hoover case.

The compressibility $\kappa$ of equations \ref{eq:LPRa} to \ref{eq:LPRd} in the
zero friction limit determines the uniformity of the background system
in which we wish to perform Langevin Dynamics. For this system

\begin{eqnarray}
\label{eq:kappaLPR}
\kappa &=& \sum_{i=1}^{N}\nabla_{\mathbf{r}_{i}}\cdot\mathbf{r}_{i}
       + \sum_{i=1}^{N}\nabla_{\mathbf{p}_{i}}\cdot\mathbf{p}_{i} \nonumber \\
       &+& \sum_{\alpha=1}^{d}\sum_{\beta=1}^{d}\frac{\partial\left(\dot{\mathbf{h}}\right)_{\alpha,\beta}}
          {\partial \left(\mathbf{h}\right)_{\alpha,\beta}}
       + \sum_{\alpha=1}^{d}\sum_{\beta=1}^{d}\frac{\partial\left(\dot{\mathbf{p}}_{g}\right)_{\alpha,\beta}}
          {\partial \left(\mathbf{p}_{g}\right)_{\alpha,\beta}},
\end{eqnarray}
which can be simplified to

\begin{eqnarray}
       \kappa &=& (d-1)\mathrm{Tr}\left[\mathbf{p}_{g}\right]/W
       - d\mathrm{Tr}\left[\mathbf{p}_{g}\right]/W
       + d\mathrm{Tr}\left[\mathbf{p}_{g}\right]/W + 0  \nonumber\\
       &=& (d-1)\mathrm{Tr}\left[\mathbf{p}_{g}\right]/W.
\end{eqnarray}

We then identify the scalar strain rate $\dot{.psilon}$ as $\mathrm{Tr}\left[\mathbf{p}_{g}\right]/Wd$, and
can therefore construct the Jacobian of the co-ordinate transform which takes the system from 
$t=^{\prime}0$ to $t^{\prime}=t$ as

\begin{equation}
\label{eq:PRJacobian}
J\left(t;0\right) = \exp\left(\int_{0}^{t}d(d-1)\dot{.psilon}\, dt^{\prime}\right),
\end{equation}
and hence the phase space metric is

\begin{equation}
\label{eq:LPRmetric}
\sqrt{g\left(t;0\right)}=\frac{V_{ref}^{d-1}}{\mathcal{V}^{d-1}} \propto
 \frac{1}{\mathrm{det}\left(\mathbf{h}\right)^{d-1}}, 
\end{equation}
where $V_{ref}$ is the volume of the cell to which the strain is referenced. We therefore
expect the correctly thermostatted probability distribution in the extended
phase space to be

\begin{equation}
\label{eq:PRPhase}
\rho\left(\mathbf{r}^{N},\mathbf{p}^{N},\mathbf{h},\mathbf{p}_{g}
\right)\propto
\exp{\left[-H^{\prime}/k_{B}T\right]}
\det\left[\mathbf{h}\right]^{1-d}.
\end{equation}

Integration of this over the $d^2$ components of the strain momentum
tensor $\mathbf{p}_{g}$ again yields a constant, yielding

\begin{equation}
\label{eq:PRcorrectPhase}
\rho\left(\mathbf{r}^{N},\mathbf{p}^{N},\mathbf{h}
\right)\propto
\exp{\left[-(\mathcal{H}+
P_{ext}\det\left[\mathbf{h}\right])/k_{B}T\right]}
\det\left[\mathbf{h}\right]^{1-d},
\end{equation}
which has previously been identified \cite{MTKCorrection}
as the correct phase space distribution for the isobaric-isothermal
ensemble with a fully flexible cell. We can therefore conclude that
the Parrinello-Rahman Langevin dynamics scheme will be capable of
correctly sampling this ensemble if equation \ref{eq:PRPhase} is
a solution of the following Fokker-Planck equation obtained from
equations \ref{eq:LPRa} to \ref{eq:LPRd}.

\begin{widetext}

\begin{eqnarray}
\label{eq:PRFokkerPlanck}
\frac{\partial\rho}{\partial{t}}&+&
\sum_{i=1}^{N}\left\{
\left(\frac{\mathbf{p}_{i}}{m_{i}} + \frac{\mathbf{p}_{g}}{W_{g}}\mathbf{r}_{i}\right)
\cdot\nabla_{\mathbf{r}_{i}}
\rho
+\left[
  \mathbf{f}_{i}-
  \frac{\mathbf{p}_{g}}{W_{g}}\mathbf{p}_{i}
  -\left(\frac{1}{N_{f}}\right)
  \frac{\mathrm{Tr}\left[\mathbf{p}_{g}
      \right]}
{W_{g}}\mathbf{p}_{i}\right]
\cdot\nabla_{\mathbf{p}_{i}}
\rho
\right\} \nonumber \\ 
&+&\sum_{\alpha,\beta}\left\{
\left[
(\mathbf{X}_{\alpha\beta}-P_{ext}\delta_{\alpha\beta})\det\left[\mathbf{h}\right]
+\frac{1}{N_{f}}\sum_{i=1}^{N}\frac{\mathbf{p}^{2}_{i}}{m_{i}}
\right]\frac{\partial \rho}{\partial (p_{g})_{\alpha\beta}}
+\left(\frac{\mathbf{p}_{g}\mathbf{h}}{W_{g}}\right)_{\alpha\beta}
\frac{\partial \rho}{\partial (h)_{\alpha\beta}}
\right\} \nonumber \\
&=& \gamma\sum_{i=1}^{N}\nabla_{\mathbf{p}_{i}}
\cdot\left[\mathbf{p}_{i}\rho
+mk_{B}T\nabla_{\mathbf{p}_{i}}
\rho\right]
+\gamma_{p}\sum_{\alpha,\beta}
\frac{\partial}{\partial (p_{g})_{\alpha\beta}}
\left[
(p_{g})_{\alpha\beta} \rho
+W_{g}k_{B}T \frac{\partial \rho}{\partial (p_{g})_{\alpha\beta}}\right]. 
\end{eqnarray}

\end{widetext}

Analysis of the LHS yields zero. This is equivalent
to the statement that the extended phase space obeys the generalised Louiville
theorem in the limit of zero friction. The background system in which the 
Langevin dynamics simulation is conducted therefore has a constant, but non-uniform
probability density which conserves equation \ref{eq:PRHPrime}.

This introduces the possibility that the Stokes-Einstein relation which we have
used to balance the diffusion and friction terms in our Langevin equations
may be invalid. When balancing these two terms (equation \ref{eq:balance}) we 
have implicitly assumed that the background probability density is flat, i.e. no
background probability gradients. Fortunately all such gradients in the Parrinello-Rahman system
are perpendicular to the particle momentum axes and hence the first term on the 
RHS of equation \ref{eq:PRFokkerPlanck} represents balanced Langevin dynamics and
is zero.

The second term of the RHS is however not zero. The background probability gradients do effect
the balance of diffusion and friction for the Langevin equation in the cell momentum. In the
case where symmetrised pressure and cell buffetting tensors have been used to eliminate cell
rotations, the imbalance is small and proportional to $\gamma_{p}$. Purists may therefore
which to remove the cell thermostat. In practise, the effect makes little difference in 
real simulations and so it is possible to use a finite $\gamma_{p}$ to aid equilibration.
Any benefits this introduces are however, likely to be small.


\section{Examples}

\subsection{Lennard-Jones Argon}
\label{sec:Examples}

In this section we will follow through the s.ps of
conducting an isotropic constant pressure Langevin dynamics simulation
for high pressure liquid argon modelled using the
familiar Lennard-Jones potential. In particular we
will investigate the temperature and density fluctuations of this system
at 80\,K, 3.36\,GPa. This corresponds
to a total energy of approximately 0.105\,Hartree for 256 atoms occupying
a cubic cell of dimension 19.05\,\AA. 

We will also simulate a solid argon system using the Parrinello-Rahman
based scheme and again investigate the quality of the fluctuations.

\begin{figure}
\includegraphics*[scale=0.5]{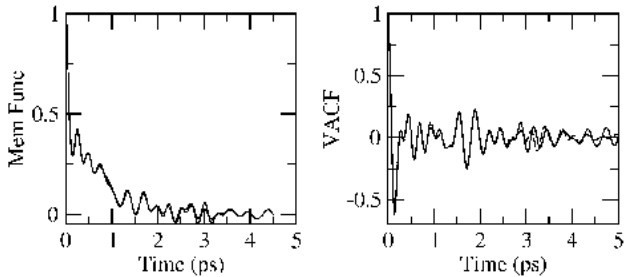}
\caption{NVE (solid line) and NVT Nos\'e-Hoover (dashed line)
VACF and memory function for Lennard-Jones argon at 
80\,K 3.36\,GPa. These are computed from 256 atom trajectories
using the autoregressive
model of Kneller and Hinsen \cite{KnellerH01} (order 150)
as implemented in the nMoldyn program \cite{RogMHK03}.}
\label{fig:VacFMem}
\end{figure}

\begin{figure}
\includegraphics*[scale=0.5]{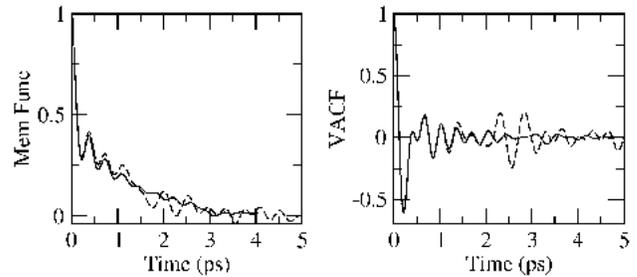}
\caption{VACF and memory function for 4000 (solid-line) and
256 (dashed-line) Lennard-Jones argon NVT Nos\'e-Hoover
simulations at 80\,K and
equal densities equivalent to 3.36\,GPa.}
\label{fig:HugeNVTSim}
\end{figure}

\subsubsection{Step 1}

First we obtain a memory function
relevant to this state point by conducting an NVE simulation
(energy 0.105 Hartree, volume (19.05\,\AA)$^{3}$).

The resulting VACF and memory functions are shown in figure
3. We also plot the equivalent functions
calculated from a Nos\'e-Hoover NVT simulation to ensure our
calculated value of $\gamma$ will be appropriate for
a system coupled to a heat bath. As can be seen the two 
memory functions are near identical in this case. 
Numerical integration under the memory function reveals an
appropriate value for $\gamma$ of $2.41 \times 10^{-3}$\,rad fs$^{-1}$.

Note that in order to ensure that the value of $\gamma$ chosen
is truly representative of a bulk liquid system, we must ensure that
the simulation used in its identification covers a length
scale larger than that of any spacial correlations. For the purposes
of this example we therefore also conduct a simulation using
4000 atoms and compute a memory function. The result is shown in
figure 4 and indicates that our value of
$\gamma$ is suitable.

\subsubsection{Step 2}

We now conduct an NVT Langevin dynamics simulation using the
optimally identified value of $\gamma$ as described in section
\ref{sec:NVTLangevin}. This allows a temperature spectrum
to be calculated, providing a criterion for choosing
a barostatic timescale. A run of 1638400 time-s.ps 
was performed in this case to provide a suitably
illustrative example, however much shorter runs can 
be used in practise.  As this system is liquid, we do not
expect any significant features in the spectrum as confirmed
in figure 5, and we therefore choose the
barostat to operate in low frequency region below the influence
of the thermostat i.e. $\omega_{b} = 2\times 10^{-4}$\,rad fs$^{-1}$.

\begin{figure}
\includegraphics*[scale=0.5]{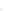}.
\caption{NVT (solid line) and NPT (dashed line)
Langevin dynamics temperature Fourier transforms for the
Lennard-Jones system at 80\,K 3.36\,GPa. The barostat frequency is
chosen so as to only affect the low frequency components introduced
by the thermostat, and not the higher frequency components characteristic
of the liquid itself.}
\label{fig:NVTFFT}
\end{figure}

\subsubsection{Step 3}

With suitable values of $\omega_{p}$ and $\omega_{b}$ identified,
a reliable NPT simulation can be conducted. We have now also
identified the timescales associated with the problem and can safely
increase the calculation time-step from 1.2 to 9.6 fs. Using these
parameters the system is equilibrated for 50,000 $\Delta t$  and
then sampled every 10 time-s.ps for a further 500,000. The
resulting temperature and volume distributions are shown
in figure 6.

\begin{figure}
\includegraphics*[scale=0.5]{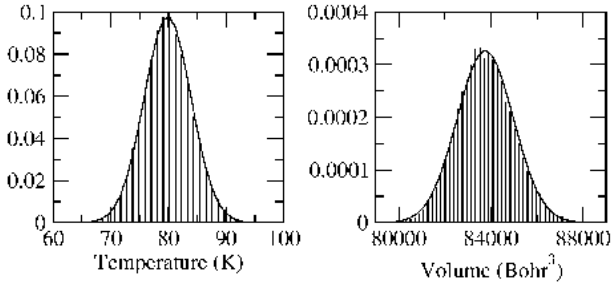}
\caption{Temperature and volume sample distributions calculated
from Langevin-Hoover simulation compared
to exact canonical cases for the Lennard-Jones argon system
at 80\,K 3.36\,GPa. The compressibility required to calculate the
ideal volume distribution was obtained from an earlier Nos\'e-Hoover
based NPT calculation.}
\label{fig:ex1dists}
\end{figure}






\subsubsection{Solid Argon with Full Cell Fluctuations}

To demonstrate the ability of the Langevin-Parrinello-Rahman (LPR) scheme to
produce equally accurate temperature and volume sample distributions, we
have also conducted simulations at 20\,K, 67.6\,MPa where argon is a
stable solid. Again memory functions were calculated from appropriate
NVE and NVT simulations, and an appropriate frequency for the barostat
was identified from the temperature spectrum as above. This 
process revealed suitable parameters of $\omega_{p} = 1.33 \times 10^{-3}$\,rad 
fs$^{-1}$ and $\omega_{w} = 4.17 \times 10^{-5}$\,rad 
fs$^{-1}$.

These values were used in a simulation of a cubic cell of FCC argon containing 
256 atoms. This system was equilibrated for 20,000 $\Delta$t and then
sampled every 10 $\Delta$t for a further 750,000. A time-step of 9.6 fs was used.
The resulting evolution of cell vectors in shown in figure 7, and
the calculated distribution of temperature and volume samples is shown in 
figure 8. As with the liquid simulations using the
isotropic algorithm, the distributions are of a very high quality.

\begin{figure}
\includegraphics*[scale=0.5]{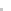}
\caption{Evolution of simulation cell during the 256 atom cubic
cell LPR simulation of argon at 20\,K 67.6\,MPa.}
\label{fig:arprcell}
\end{figure}

\begin{figure}[b]
\includegraphics*[scale=0.5]{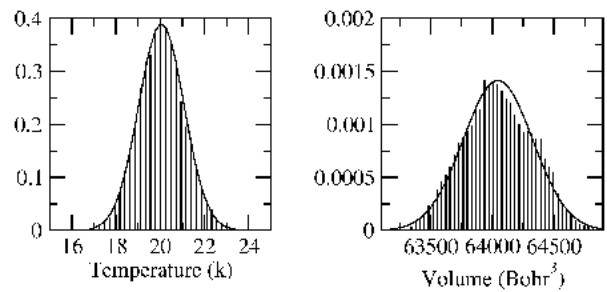}
\caption{Temperature and volume sample distributions of solid
FCC argon at 20\,K 67.6\,MPa calculated using the LPR scheme. Exact
canonical distributions are shown as solid lines.}
\label{fig:arprdists}
\end{figure}

\subsection{\emph{Ab-initio} Silicon}

In this section we will use the LPR scheme for the more realistic application
of simulating silicon. We will first use the semi-empirical potential of Tersoff
\cite{Tersoff89} to obtain an approximate memory function for crystalline
silicon at room temperature and pressure. We will then use this
memory function to choose parameters for a smaller silicon system 
using the LPR implementation in the \emph{ab-initio} plane wave DFT code
CASTEP \cite{Castep}. 

\subsubsection{Step 1}

Again we perform both NVE and NVT runs with 216 atoms at the temperature
and pressure of interest (using the Tersoff potential) in order to obtain a memory
function characteristic of the system in question. These
are shown in figure 9. A time-step of $\Delta t = $\,2.0 fs
was used for a total run of 20,000 $\Delta t$ sampling
every 4 time-s.ps after an initial equilibration period of
10,000 $\Delta t$. The system was initialised with the expected
equilibrium diamond structure.

Integration over the memory function yields a suitable thermostat frequency
of $3.05 \times 10^{-4}$\,rad fs$^{-1}$. This corresponds to
a thermostat period of 20.81\,ps.

\begin{figure}
\includegraphics*[scale=0.5]{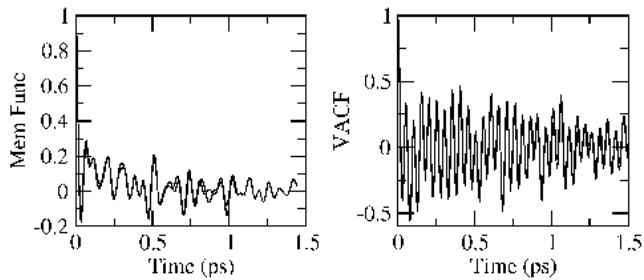}
\caption{NVE (solid line) and NVT (dashed line)
VACF and memory function for the 216 silicon atom system modelled
with the Tersoff potential at room temperature and atmospheric pressure.}
\label{fig:SiVacFMem}
\end{figure}

For interest, we also compute a memory function for a Nos\'e-Hoover
NVT simulation at the same state point using CASTEP. This was calculated for an eight atom cell
using a cut-off energy of 160\,eV and a time-step of 8 fs. The ultra-soft pseudo-potential method was used.
The Brillouin zone was sampled at 4 k-points using the Monkhorst-Pack method, and the exchange
and correlation functional was approximated using the LDA. The result is shown 
in figure 10. Although we expect this memory function to be subject to 
significant finite-size effects, the 
thermostat period obtained from this memory function is 16.95\,ps, similar to that obtained from
the larger system.

\begin{figure}
\includegraphics*[scale=0.5]{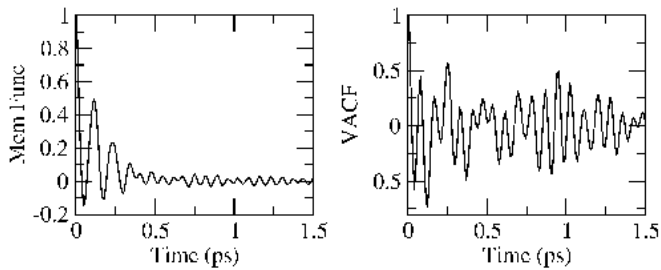}
\caption{Memory function and VACF calculated from the  8-atom
\emph{ab-initio} simulation of silicon at room temperature and pressure. These
are expected to exhibit severe finite size effects, but are similar to those obtained from
the Tersoff potential in figure 9.}
\label{fig:castepmem}
\end{figure}

\subsubsection{Step 2}

An NVT Langevin dynamics simulation at the required temperature and pressure
(using the Tersoff potential)
of 163,840 $\Delta t$
sampled every 10 $\Delta t$
yields the temperature 
spectrum shown in figure 11. For this
example we use a value of $\omega_{b}$
set to $3.5\times 10^{-3}$\,rad fs$^{-1}$ which is
well separated from the dominant frequencies and
a few times smaller than that of the thermostat.
The temperature profile for a NPT run at this value of 
$\omega_{b}$ is also shown in figure  11, indicating
that the choice is appropriate.

\begin{figure}[b]
\includegraphics*[scale=0.5]{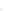}
\caption{NVT (solid line) and NPT (dashed line)
Langevin dynamics temperature Fourier transforms for the 216 atom
silicon system modelled using the Tersoff potential. The barostat frequency
has been chosen to avoid disruption of the natural components and hence the
two profiles are similar.}
\label{fig:SiTFFT}
\end{figure}

\subsubsection{Step 3}

With appropriate thermostat and barostat timescales obtained
from the classical potential, we now perform the \emph{ab-initio}
LPR simulation. Again we use 4 k-points and a plane wave cut-off
of 160\,eV. The system was initialised in the ideal BC8 structure
(cell parameter 5.75\,\AA) with random velocities, and equilibrated
for 8,500 $\Delta$ t. The system was then sampled at a further 
6,200 
s.ps. This represents
a relatively simple simulation, requiring modest CPU time.

The resulting cell evolution is shown in figure 12. We also
calculate the distribution of temperature and volume samples as shown in 
figure 13. For such a small simulation, we do not expect
exact fits to canonical distribution functions, however the plots indicate
convergence toward the ideal results as demonstrated for argon above.
  
\begin{figure}
\includegraphics*[scale=0.45]{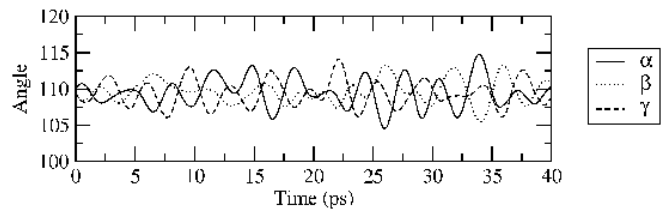}
\caption{Evolution of cell angles during the 8 atom \emph{ab-intio}
LPR simulation.}
\label{fig:si8cell}
\end{figure}

\begin{figure}
\includegraphics*[scale=0.5]{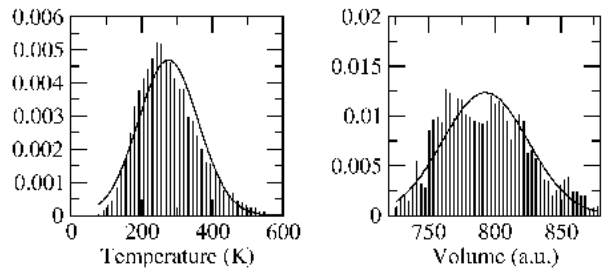}
\caption{Distribution of temperature and volume samples during the 8 atom \emph{ab-intio}
LPR simulation. The solid lines are the calculated canonical bulk values for comparison.
although it
should be noted that both distributions \emph{should} differ from the bulk cases which
are derived for large N.
}
\label{fig:si8dists}
\end{figure}

\section{Discussion}

The results presented in the previous section have demonstrated that Langevin-Hoover
and Langevin-Parrinello-Rahman schemes are capable of producing canonical fluctuations
from which ensemble specific thermodynamic quantities can be calculated. Clearly
the choice of parameters is important. 

In this paper we have calculated optimal damping co-efficients (and hence thermostat
frequencies $\omega_{p}$) for the stochastic
component of the dynamics by computing a memory function for our example simulations. While this has provided
optimal sampling efficiency, it has also complicated the choice of the barostat
frequency $\omega_{b}$. We have stated earlier that the barostat frequency
should be smaller than, but close to that of the simulated particles. This allows the 
particle motions to modulate the volume oscillation and hence generate canonical
fluctuations. Too low a barostat frequency, 
and its motion becomes decoupled from the rest of the
simulation, resulting in an undesirable slow harmonic oscillation of the volume. 
We also require the thermostat frequency $\omega_{p}$
to be significantly higher than the barostat frequency $\omega_{b}$. This prevents disruption of the 
thermostatic process by the cell motion.

As we have chosen our Langevin friction co-efficient $\gamma$ to be characteristic
of the particle motions, $\omega_{p}$ will be close to the particle frequency $\omega$. We also require
$\omega_{b}$ to be close to $\omega$ but well separated from $\omega_{p}$. The choice of
an optimal friction co-efficient therefore introduces an apparent conflict in the choice
of barostat parameter. We have shown above in section \ref{subsec:choice} that this issue can be overcome
by use of the temperature spectrum in carefully choosing the barostat parameter.

It is important to note that this is not always the case in practise.
The optimal friction co-efficient identified from the memory function represents
the boundary between performing molecular dynamics with a  `Langevin thermostat' (i.e. low $\omega_{p}$ with similar
correlation times and sampling efficiencies as a Nos\'e-style thermostat for 
chaotic systems) and performing a `true Langevin dynamics' simulation where the stochastic components
dominate and are guided by the particle interaction forces (high $\omega_{p}$). For the purposes of
sampling an isothermal ensemble, a small conservative choice of $\omega_{p}$
will also produce canonical temperature fluctuations, but 
on a longer timescale. The value calculated from the memory function therefore 
represents the best compromise between efficiency and preservation of accuracy
in short-term dynamics.

As a conservative rule of thumb, $\omega_{p}$ can be set a few tens of
times smaller than the particle frequency $\omega$, and $\omega_{b}$ can 
be chosen a few times smaller again. This allows canonical simulations in either the isotropic or
fully flexible cell NPT ensemble to be generated with ease, albeit with
a reduced sampling efficiency due to weaker modulation of the volume
oscillation. A compromise must therefore be reached between
the computational expense of computing optimal parameters, and statistical
efficiency in the simulation itself.

This stochastic approach
bypasses any concerns about possible `hidden' conservation laws \cite{TuckermanLCM01}
in Nos\'e style schemes which can incorrectly restrict trajectories and lead to
incorrect ensemble averages. This effect in a simple harmonic potential
is well documented (although still not entirely understood) and is generally
thought to be ignorable in practical simulations. Often a chain
of thermostats is used to break any unphysical conservation laws, however 
this process takes place on a longer timescale than
in a stochastic thermostatting scheme - an important consideration in \emph{ab-initio}
dynamics.

Solids at low temperature are essentially harmonic, as are the forces used
to `connect' different realisations in imaginary time during path-integral
molecular dynamics simulations. We therefore suggest that the isobaric-isothermal
sampling schemes presented here are a desirable alternative
to traditional deterministic schemes for many applications.

\section{Conclusions}

We have shown that performing Langevin dynamics within constant
pressure extended systems is a valid method for simulating
the equilibrium isobaric-isothermal ensemble. An analysis of the phase
space in these simulations has been presented, which shows that
the correct distribution of probability is generated,
provided the non-Hamiltonian nature of the extended system is
accounted for.

We have derived a suitable integration scheme, analogous to the
Velocity-Verlet scheme, which allows reversible integration
of trajectories in the zero friction limit, and eliminates
long term drift of the conserved quantity in the extended
system when the stochastic component is introduced.

Suitable distributions of 
temperature and volume samples can be achieved provided a little
care is taken in choosing parameters. Methods for identification
of optimal values of these parameters have been presented, We
have also discussed less stringent criteria for choosing parameters
in the case where computing a memory function or an accurate NVT
temperature spectrum is computationally prohibitive. 

We consider this scheme to be a desirable alternative to extended
system NPT schemes with Nos\'e-style thermostats, in particular for small or approximately harmonic
systems, where deterministic schemes can suffer from unexpected
coupling to the particle sub-system leading to incorrect trajectories, and
consequent thermodynamics averages.

\section{Acknowledgements}

Financial support for this research has been partly provided by an EPSRC studentship. 
Computational facilities were provided under EPSRC grant R47769.

\bibliography{refs}

\end{document}